# Statistical Laws in the Income of Japanese Companies

Takayuki Mizuno[1], Makoto Katori[1], Hideki Takayasu[2] and Misako Takayasu[3]

[1]Department of Physics, Faculty of Science and Engineering, Chuo University, Kasuga, Bunkyo-ku, Tokyo 112-8551, Japan
[2]Sony Computer Science Laboratories Inc., 3-14-13 Higashigotanda, Shinagawa-ku, Tokyo 141-0022, Japan
[3]Department of Complex Systems, Future University-Hakodate, 116-2 Kameda-Nakano-cho, Hakodate, Hokkaido 041-8655, Japan

**Summary.** Following the work of Okuyama, Takayasu and Takayasu [Okuyama, Takayasu and Takayasu 1999] we analyze huge databases of Japanese companies' financial figures and confirm that the Zipf's law, a power law distribution with the exponent -1, has been maintained over 30 years in the income distribution of Japanese companies with very high precision. Similar power laws are found not only in income distribution of company's income, but also in the distributions of capital, sales and number of employees.

From the data we find an important time evolutionary property that the growth rate of income is approximately independent of the value of income, namely, small companies and large ones have similar statistical chances of growth. This observational fact suggests the applicability of the theory of multiplicative stochastic processes developed in statistical physics. We introduce a discrete version of Langevin equation with additive and multiplicative noises as a simple time evolution model of company's income. We test the validity of the Takayasu-Sato-Takayasu condition [Takayasu, Sato and Takayasu 1997] for having an asymptotic power law distribution as a unique statistically steady solution. Directly estimated power law exponents and theoretically evaluated ones are compared resulting a reasonable fit by introducing a normalization to reduce the effect of gross economic change.

**Key words.** Zipf's law, Income distribution, Random multiplicative process, Discrete Langevin equation, Takayasu-Sato-Takayasu condition

## 1. Income Distributions of Japanese Companies

The databases we analyze in this paper are lists of large Japanese companies' annual incomes reported to the tax offices, which have been published by a publisher *Diamond*, Tokyo, since 1970. Here, the income is roughly defined by the total incoming cash flows minus outgoing ones before tax, and the tax rate for private companies in Japan is about 50% uniformly.

Let $I_i(t)$ be the annual income of the $i$-th company at year $t$ with $i = 1,2,\cdots,N$. We define the probability density of income distribution $P_t(I)$ as $P_t(k\Delta I)\Delta I \equiv |\{i : I_i(t) \in [k\Delta I,(k+1)\Delta I]\}|/N$ for $k = \{0,1,2,\cdots\}$, where $\Delta I$ is the bin size for counting income, and $|A|$ denotes the number of elements of the set $A$. The cumulative distribution is defined as

$$P_t(\geq k\Delta I) = \sum_{n \geq k} P_t(n\Delta I).  \qquad (1.1)$$

Fig.1 is the log-log plot of the cumulative distribution $P_t(\geq I)$ for about 80 thousand companies whose incomes exceed 40 million yen in the year of $t = 1998$. Clearly we can confirm the power-law distribution with the exponent very close to $-1$, i.e., so-called the Zipf's law.

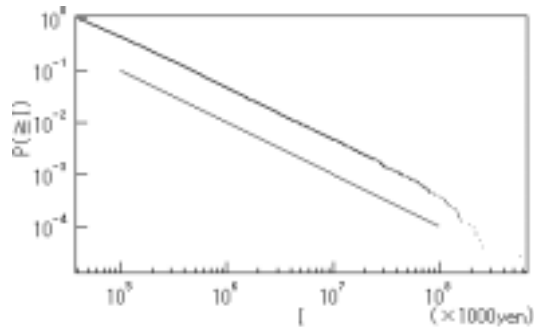

**Fig. 1.** Cumulative distribution of company's incomes of Japan in 1998 in log-log scale. The straight line represents the slope of the power-law with exponent –1.

Next, we investigate the income distributions of the top 1,000 Japanese companies for 30 years from 1970 to 1999 as shown in Fig. 2. All of the income distributions are clearly parallel to the line having the slope –1 in the log-log plot, namely, the Zipf's law holds for 30 years.

It should be noted that the income distributions are shifted year by year. Fig.3a shows the shift of income distributions to the right during the period from 1983 to 1989, so-called the *bubble period,* however, after the collapse of the *bubble* the distribution suddenly shifted to the left in 1991. From this result we may say that the shift of the income distribution can be regarded as a good indicator of gross economic condition. Fig. 3b shows that the income distributions are shifted to the left in 1991-1995 and in 1997 they turned the shift direction to the right. This result may imply that Japanese economic condition is recovering since 1995.

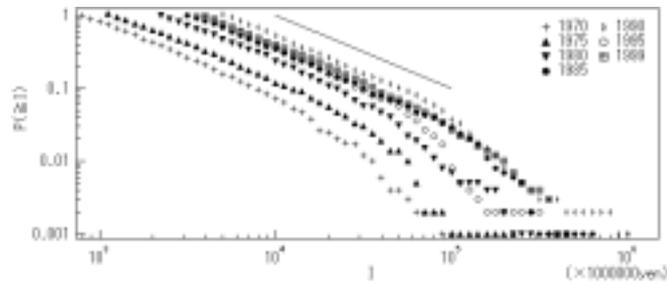

**Fig. 2.** Cumulative distribution of company's incomes of Japan from 1970 in log-log plot. Data of the top 1,000 companies are plotted in every 5 years. The straight line indicates the power-law with exponent –1.

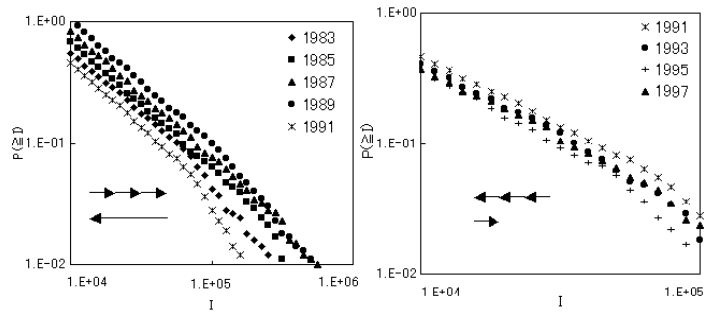

**Fig. 3** Shifts of the income distributions. 3a: During the period of so-called *the bubble economic condition* (1983 to 1990) the incomes grow larger shown by the shift to the right. After the collapse of the bubble the incomes shifts to the left in 1991. 3b: Shifts of the income distributions in recent years.

## 2. Power-law Distributions of Other Quantities

Power-law distributions are found not only in income but also in other economical data of companies. The data of capital and sales of large Japanese companies are also published by the same publisher *Diamond* and we analyze each data in the same way by observing the cumulative probability distributions. The following figures (Fig. 4a – 4b) show these distributions demonstrating the validity of the same Zipf's law.

We can also investigate the distribution of number of company's employees from 1969 to 1996 using the database of Statistics Bureau, Management and Coordination Agency, Japan (Fig. 4c). The power law behavior close to the Zipf's law is again observed also in the number of company's employees. From these results we believe that the power-law tail of exponent close to –1 is a kind of universal property characterizing the statistics of Japanese company's financial sizes.

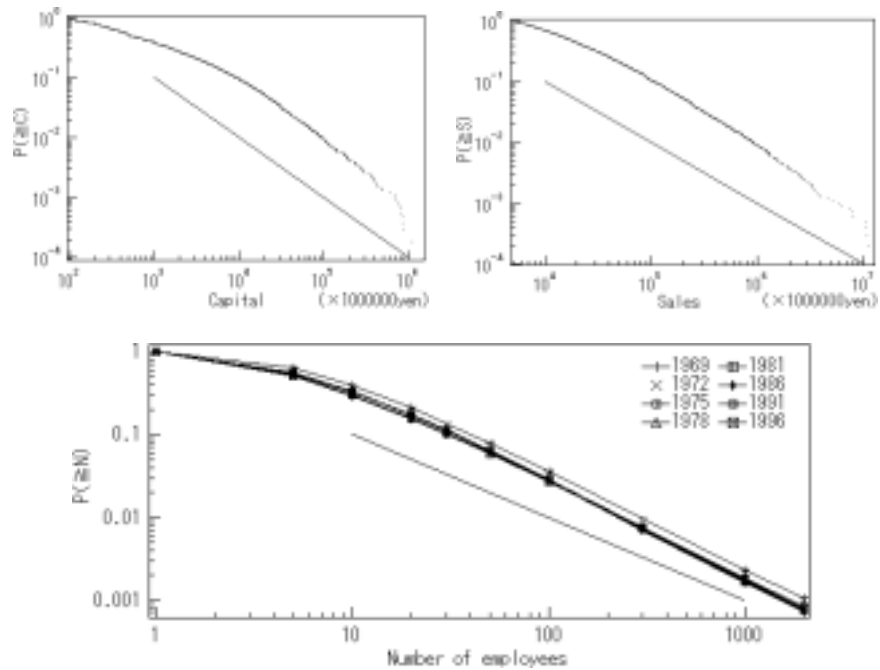

**Fig. 4a,b,c. a** : Upper left figure shows the cumulative distributions of the capital. The data points contain about 10,000 companies with the capital of 100 million yen or more. **b** : Upper right figure shows the cumulative distributions of the sales. The number of companies is about 10,000 companies with the annual sales of 5 billion yen or more. **c** : Lower figure shows the cumulative distributions of number of employees. The straight line indicates the power-law with exponent –1 in each figure.

## 3. Dependence of Income Distribution on Job Categories

So far, we have analyzed the data as a whole, here we analyze the income distributions in more detail taking into account the effect of job categories. The companies are classified into 43 job categories in the database. As represented in Fig.5 the Zipf's law holds for income distributions only in a few job categories and there are many examples showing power law behaviors but the exponents clearly deviate from –1. For example, the income distribution of Electrical Product companies has the exponent, –0.72, significantly different from -1.

There are some cases in which the income distributions do not show power laws. For example, in the case of banks the plots are apparently not on a straight line in the log-log plots. The Japanese governmental policy of protecting banks from bankrupt might be responsible for this special case.

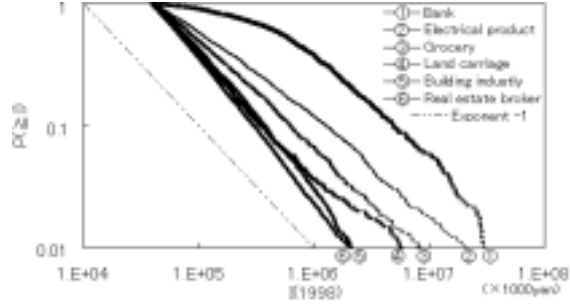

**Fig. 5.** Income distributions of several job categories. The straight line indicates the power law with exponent –1.

**Table 1.** Examples of Exponents of income distribution of each job category.

| Job Category | Exponent |
| --- | --- |
| Electrical Product | –0.78 |
| Grocery | –0.87 |
| Land carriage | –0.92 |
| Building industry | –1.12 |
| Real estate broker | –1.25 |
| Bank | not power law |

## 4. Randomly Multiplicative Process

In order to estimate the underlying dynamics of income fluctuations we study the ratio of incomes of two consecutive years, $R_i \equiv I_i(t+1)/I_i(t)$. We fix the bin size of the ratio, $\Delta R$, and observe the probability density defined as

$$P_t(k\Delta I, m\Delta R)\Delta I \Delta R \equiv \left|\{I_i(t) \in [k\Delta I, (k+1)\Delta I], \ R_i \in [m\Delta R, (m+1)\Delta R]\}\right|/N \quad (3.1)$$

for $k, m \in \mathbf{Z}$. The conditional probability density of $R_i(t)$ for given $I_i(t)$ is given by $P_t(R|I) \equiv P_t(I,R)/P_t(I)$. We analyze the data $\{(I_i(t), I_i(t+1)) : i = 1,\cdots,N\}$ for $N = 4,500$ companies at $t = 1997$, and $P_t(R|I)$ is evaluated as shown in Fig. 6. We find that the dependence of functional from of $P_t(R|I)$ on $I$ is very weak. This implies that the distribution function $P_t(I,R)$ is simply given as

$$P_t(I, R) \cong P_t(I) \cdot P_t(R). \quad (3.2)$$

Moreover, by a numerical fitting we find that $P_t(R)$ is well approximated by the following function,

$$P_t(R) = \begin{cases} c(R/\langle R \rangle)^{\tau_-} & \text{for} \quad 0.08 < R < \langle R \rangle \\ c(R/\langle R \rangle)^{-\tau_+} & \text{for} \quad \langle R \rangle \leq R < 5 \end{cases}, \qquad (3.3)$$

as shown in Fig. 6, where $\langle R \rangle = 0.89, c = 0.08, \tau_- \cong 1$ and $\tau_+ \cong 3.5$.

This observation result implies that the time-evolution of the income is well approximated by a simple randomly multiplicative process such as

$$I_i(t+1) \cong R \cdot I_i(t), \qquad (3.4)$$

where the factor $R$ is a random number independent of the value of income $I_i(t)$.

Stanley *et al.* [1996] and Lee *et al.* [1998] analyzed a similar conditional probability densities for the growth rates of sales of USA companies and the GDP of countries, and reported that the growth rates generally follow symmetric exponential distributions. It should be noted that for sales and GDP these conditional probabilities show strong dependence on the company's sizes through their standard deviations, however, for the growth rates of incomes there seems to be no such size dependence.

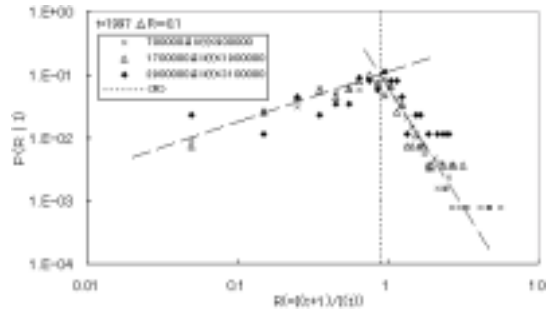

**Fig. 6.** Conditional probability density of growth ratio , $R$.

## 5. Takayasu-Sato-Takayasu (TST) Condition

As a basic start point of establishing the numerical model of company's incomes we review the following theorem for the discrete version of the linear Langevin equation for a stochastic variable $x(t)$ in time $t$ [Takayasu, Sato and Takayasu 1997],

$$x(t+1) = b(t)x(t) + f(t), \qquad (4.1)$$

where $b(t)$ and $f(t)$ are random variables with the probability density functions $W(b)$ and $U(f)$, respectively. Takayasu, Sato and Takayasu (TST) exactly calculated the characteristic function of $x(t)$, $\langle e^{i\rho x(t)} \rangle$, where $\langle \cdots \rangle$ is the average over realizations of stochastic process. They proved the following general statement.

**Theorem (Takayasu, Sato and Takayasu 1997).** *Assume that $W(b)$ and $U(f)$ are time-independent and $U(f)$ is a continuous function of $f$. Let $G(\beta) = \langle b^\beta \rangle$. If and only if*

$$\lim_{\beta \to \infty} \frac{dG(\beta)}{d\beta} = \langle \log b \rangle < 0 \quad \text{and} \quad G(2) = \langle b^2 \rangle > 1, \quad (4.2)$$

*the equation (4,1) has a unique stable steady solution such that it has power-law tails in the cumulative distribution:*

$$P(\geq |x|) \quad |x|^{-\beta}. \quad (4.3)$$

*The exponent $\beta$ of this power-law tail is determined by the following equation*

$$G(\beta) = \langle b^\beta \rangle = 1. \quad (4.4)$$

Here, we call the condition (4.2) as the TST condition. The equation (4.4) is called the TST equation.

## 6. Data fitting of the Langevin model

We analyze the data of incomes of two consecutive years, $\{(I_i(t), I_i(t+1) : i = 1, 2, \cdots, N\}$ for $N = 4500$ in the years $t$=1996 and $t$=1997. Here, we test the TST conditions (4.2) and the TST equation (4.4) for real data of income and try to explain the power laws observed in the preceding sections. We also discuss the relation between the shift of the distribution and the deviation of the slopes of income distributions in each job category.

As already seen in Figs. 2 and 3 the distributions of income are shifted depending on the gross economic conditions. In order to apply the stationary statistics of the Langevin type equation we need to normalize the incomes so that they can be treated as statistically stationary independent of the gross economic condition. As the GDP and the assessed land value are conventionally used as indicators of gross economic condition, we observe the change of GDP and compare it with the incomes of the 350th company and 1000th company as shown in Fig.7. We can find that the GDP is continuously growing up, however, there are bending points in the incomes of the 350th and 1000th companies. This result implies that the normalization of incomes using the GDP is not appropriate. We also check the normalization by the assessed land value, however, we meet the same difficulty as the case of the GDP, so the assessed land value is not good for the normalization of incomes.

As there is no known macro-value that can be used for the normalization, here we introduce a simple normalization of the incomes of each year by the amount of the income of the 1000th company in the ranking of the year;

$$X_i(t) \equiv I_i(t) / I_{1000}(t).  \qquad (5.1)$$

We define the ratio of two consecutive years as

$$R_i(t) \equiv X_i(t+1) / X_i(t), \qquad (5.2)$$

where the suffix $i$ specifies each company.
Using the normalized quantity we can observe the following relation,

$$\langle R(1996) \rangle \langle R(1997) \rangle = 1.101, \quad \langle R(1996) R(1997) \rangle = 1.103.$$

Namely, the time correlation of the normalized incomes' ratio, $R(t)$, becomes negligibly small, and we can expect that the TST condition is satisfied for the normalized incomes.

We can evaluate the value of $\beta$ in two ways; one is directly estimating the slope of the cumulative distribution of income, and the other is solving the TST equation for the normalized income $X_i(t)$ of data in 1996 and 1997. These values are listed in Table 2.

Next we test the validity of TST equation for income of job categories. We find a rather large discrepancy between $\beta$ (directly evaluated)=0.9 and $\beta$ (solution of the TST equation)=1.8 for Land Carriage Companies. This discrepancy is expected to be due to the strong time-correlation in this job category as shown in Table 3, where the autocorrelation is defined as

$$C \equiv \{\langle R(t)R(t+1) \rangle - \langle R(t) \rangle \langle R(t+1) \rangle\} / \{\langle R(t)^2 \rangle - \langle R(t) \rangle^2\}. \qquad (5.5)$$

We calculate the autocorrelations using the data from 1996 to 1998. Table 3 shows that there is almost no correlation in the whole Japanese companies, however, there are non-negligible correlations particularly in the land carriage companies. In order to take this effect into account we solve the following modified TST equation with the time interval of two years using the income data from 1996 to 1998,

$$\langle \{b(96)b(97)\}^\beta \rangle = 1. \qquad (5.6)$$

By this modification we obtained the corrected value, $\beta$ (corrected) $=1.1$, as shown in Fig. 8.

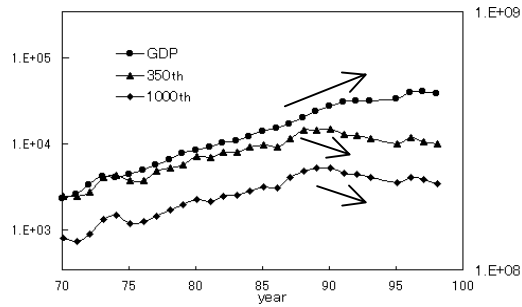

**Fig. 7.** Time dependence of the GDP and the incomes of the 350th and 1000th companies.

**Table 2.** Comparison of the value of $\beta$

| t | directly evaluated $\beta$ | solution $\beta$ of TST equation |
|---|---|---|
| 1996 | 1.01 | 0.90 |
| 1997 | 1.00 | 0.95 |

$\beta$ (directly evaluated) > $\beta$ (solution of TST equation) is reasonable, since the data with low $b(t)$ are not included systematically because the companies with incomes smaller than 40 million yen are omitted in the database.

**Table 3.** Autocorrelation of consecutive incomes

| Job Category | $C$ |
|---|---|
| All | 0.002 |
| Electrical Product | 0.020 |
| Building Industry | –0.110 |
| Real Estate Broker | –0.067 |
| Land Carriage | 0.492 |

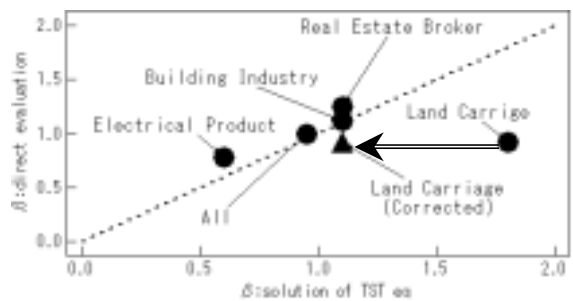

**Fig. 8.** Direct evaluation vs. solution of TST equation of the exponent $\beta$. The corrected value of Land Carriage using equation (5.6) fits better with the direct value.

## 7. Conclusions

We have confirmed that the income of the whole Japanese companies have been obeying the power law distribution with the exponent very close to –1 for 30 years. The shifts of the income plots correspond to the gross economic condition. In each job category the income distributions follow power laws in most cases and the exponents of the power-law are scattered around -1. It has been demonstrated that these exponents can be estimated theoretically from the data of growth rates using TST equation assuming that the income evolution is approximated by a simple random multiplicative process. We believe that these data analysis give a clue to clarify the underlying dynamics of complicated interactions among companies.

Potential applicability of the present theoretical formulation is expected to be very wide. Ijiri and Simon showed that the city population of USA obeys the power law distribution for a half-century [Ijiri and Simon 1977], and Aoyama et al found that incomes of individuals in Japan follow the power law with exponent –2 instead of –1[Aoyama et al 2000]. Mathematical models for these phenomena will be developed in the near future based on random multiplicative processes.

The present authors would like to thank Kenichi Tsuboi and Nobue Sakakibara in Diamond for allowing them to use the data published previously and Mitsuhiro Okazaki for useful discussions.

## References

K Okuyama, M Takaysu, H Takayasu (1999) Zipf's law in income distribution of companies. Physica A269: 125-131

H Aoyama, Y Nagahara, M P Okazaki, W Souma, H Takayasu, M Takayasu (Sept.2000) Pareto's law for income of individuals and debt of bankrupt companies. Fractale 8: 293-300 [cond-mat 10006038]

M H R Stanley, L A N Amaral, S V Buldyrev, S Havlin, H Leschhorn, P Maass, M A Salinger, H E Stanley (1996) Scaling behavior in the growth of companies. Nature 379: 804-806

Y Lee, L A N Amaral, D Canning, M Meyer, H E Stanley (1998) Universal Features in the Growth Dynamics of Complex Organizations. Phys Rev Lett 81: 3275-3278

H Takayasu, A H Sato, M Takayasu (1997) Stable Infinite Variance Fluctuations in Randomly Amplified Langevin Systems. Phys Rev Lett 79: 966-969

Yuji Ijiri, Herbert A Simon (1977) Skew distributions and the sizes of business firms. Amsterdam: North Holland Publishing